\documentclass{article} 
\usepackage{iclr2024_conference,times}
\usepackage{graphics}

\usepackage{amsmath,amsfonts,bm}

\def\eqref#1{equation~\ref{#1}}

\def\1{\bm{1}}

\DeclareMathAlphabet{\mathsfit}{\encodingdefault}{\sfdefault}{m}{sl}
\SetMathAlphabet{\mathsfit}{bold}{\encodingdefault}{\sfdefault}{bx}{n}

\usepackage{hyperref}
\usepackage{url}
\usepackage{graphicx}

\title{Density-based Neural Temporal Point Processes for Heartbeat Dynamics}

\author{Sandya Subramanian \thanks{https://www.subramanianlab.com} \\
Department of Computational Precision Health\\
University of California Berkeley\\
Berkeley, CA 94709, USA \\
\texttt{sandyas@berkeley.edu} \\
\And
Bharath Ramsundar \\
Deep Forest Sciences \\
Palo Alto, CA 94306, USA \\
\texttt{bharath@deepforestsci.com} \\
}

\iclrfinalcopy 
\begin{document}

\maketitle
\vspace*{-.2cm}
\begin{abstract}
Temporal point processes (TPPs) provide a natural mathematical framework for modeling heartbeats due to capturing underlying physiological inductive biases. In this work, we apply density-based neural TPPs to model heartbeat dynamics from 18 subjects. We adapt a goodness-of-fit framework from classical point process literature to Neural TPPs and use it to optimize hyperparameters, identify appropriate training sequence lengths to capture temporal dependencies, and demonstrate zero-shot predictive capability on heartbeat data.
\end{abstract}

\vspace*{-.1cm} \section{Introduction} \vspace*{-.1cm}
Temporal point processes (TPP) provide a powerful framework for modeling several physiological phenomena in an etiologically faithful way. One such example is heartbeats, which are known to have statistical structure governed by point processes \cite{barbieri2005point}. Recent work has introduced a framework of neural TPPs \cite{shchur2021neural} for non-physiological time series data. These models hold the potential for greater representational power for complex nonlinear dynamics and increased ease of handling big data compared to existing physiological point process models.

However, most Neural TPP methods use negative log likelihoods (NLLs) as the main metric of model evaluation. This choice is not optimal for several reasons: (1) NLLs are not comparable across sequences of different length; (2) NLLs serve as both loss and metric, which can bias evaluations; (3) NLLs are not comparable across models; and (4) most importantly, NLLs do not question the underlying generative assumptions of the model itself: whether or not the point process model in question is an appropriate choice for the data. Past work (\cite{shchur2021neural} also notes the limitations of NLL as an evaluation metric.

In this work, we introduce a goodness-of-fit (GOF) framework that draws on classical TPP theory and apply it to Neural TPPs. This framework provides a natural metric for evaluating model fit, the KS-distance (defined in Section~\ref{sec:gof}). We apply the framework to electrocardiogram (ECG) data (specifically a publicly available dataset from the Physionet collection \cite{moody2001physionet}), for which there is known statistical structure underlying the generative process. The GOF framework offers the following benefits: (1) a metric independent from model optimization, (2) a method to question underlying generative assumptions on a per dataset basis, (3) comparability across models, and (4) significance cutoffs for the KS-distance that account for sequence length. To our knowledge, this work is the first application of classical GOF frameworks to Neural TPPs and of density-based Neural TPP models to ECG data. We use the GOF framework to (1) perform systematic hyperparameter search, (2) evaluate the effect of altering training sequence lengths, and (3) evaluate zero-shot learning capability.

\section{Background and Related Work}
\label{gen_inst} \vspace*{-.1cm}
We introduce notation and mathematical details for point processes in Appendix~\ref{app:point}. 
\vspace*{-.0cm} \subsection{Goodness-of-Fit and The Time Rescaling Theorem} \label{sec:gof} \vspace*{-.0cm}
Classical temporal point process theory already has the powerful machinery of GOF frameworks for point processes. These methods have been vetted across domains and allow for evaluation of model assumptions on a per dataset basis. The foundation of these is the Time Rescaling Theorem \cite{brown2002time}, which states that any point process, defined uniquely by a time-varying conditional intensity function, can be transformed to a Poisson process with rate 1. 

To formalize, suppose we have a sequence of events $0 < t_1 < \dotsc < t_N$ that are drawn from a temporal point process with conditional intensity function $\lambda(t|\mathcal{H}_t)$. Then define a transformation 
$r_i = \int_0^{t_i} \lambda(s|H_s) ds$
that transforms the original sequence of events into sequence $0 <  r_1 < \dotsc < r_N$. Then the time-rescaling theorem states that $\mathcal{R} = \{ r_i \}$ is a Poisson process with unit rate. Given these transformed times, we can compute rescaled intervals $\rho_i := r_i - r_{i-1}$. By the time-rescaling theorem, these are independently drawn exponential random variables with mean $1$. We can perform the transformation $z_i = 1 - \exp(-\rho_i))$ and we will have that the $z_i$ are uniform random variables. 

Therefore, to test the GOF of a point process model, we can transform the empirical intervals using the time-rescaling theorem and assess whether the resulting $z_i$'s are independent, uniform random variables. To do this, we use a visualization called the Kolmogorov-Smirnov (KS) plot, which plots the quantiles of the $z_i$'s against the quantiles from a uniform distribution (Fig~\ref{fig:hero}). For a well-fitting model, the $z_i$'s should closely follow along the line $y=x$. We can quantify this by computing a KS-distance, which is the maximum vertical distance from the $y=x$ line \cite{brown2002time}. The 95\% significance cutoff for the KS-distance is defined as $1.36/\sqrt{n}$, where $n$ is the number of data points in the sequence. Another visualization technique to assess GOF is to plot pairs of subsequent intervals against each other. The original intervals from a point process are typically highly correlated, whereas the rescaled intervals ($\rho_i$'s) should be uncorrelated.   

\vspace*{-.0cm} \subsection{The Neural Temporal Point Process} \vspace*{-.1cm}
A parameterized temporal point process is governed by a set of parameters $\theta$. It is possible to start from a parameterized conditional intensity function $\lambda^*_\theta$ or from a parameterized conditional density $p^*_\theta$. In this work, we focus on the parameterized density $p_\theta^*$ following \cite{shchur2019intensity}. As we will see shortly, this choice simplifies the form of the negative log-likelihood. It also appropriately represents the known physiology of heartbeat generation (Section~\ref{sec:heart}). We can then learn $\theta$ from data by minimizing the parameterized negative log-likelihood
\begin{align}
    \theta^* &= \arg\min_\theta -\sum_i \log p^*_\theta(\tau_i)
    = \arg\min_\theta \left [ -\sum_i \log \lambda^*_\theta(t_i) + \int_0^{t_N} \lambda^*_\theta(s)ds  \right ]
\end{align} 
We parameterize the conditional density $p^*_\theta$ as a mixture of lognormal distributions:
\begin{align}
    p(\tau|w,\mu,s) &= \sum_{k=1}^K w_k \frac{1}{\tau s_k \sqrt{2\pi}} \exp \left (  -\frac{(\log \tau - \mu_k)^2}{2s_k^2}\right )
\end{align}
where $w_k, s_k, \mu_k$ are learnable parameters. Neural TPPs embed the history up to $t$, $\mathcal{H}_{t}$, into fixed embeddings $h \in \mathbf{R}^H$ where $H$ is the size of the history embedding space. In this work we use recurrent neural networks to perform this embedding $h_i = \mathrm{RNN}(t_1,\dotsc, t_{i-1})$. We also consider RNN variants like GRUs and LSTMs \cite{goodfellow2016deep}. We also allow the model to learn a sequence specific embedding $e$. The context is defined as the concatenation of the history and the embedding vectors $c = h || e$. The parameters for the density are learned functions of the context (where $V, b$ are learned from data): $w = \mathrm{softmax}(V_w c + b_w)$, $s = \exp(V_s c + b_s)$, 
$\mu = V_\mu c + b_\mu$.
\vspace*{-.0cm}\subsection{Statistical Models of Heartbeats}
\label{sec:heart} \vspace*{-.1cm}
The ECG measures electrical activity of the heart. The most notable morphology of the ECG is the R complex or R peak, which denotes the heartbeat. The intervals in between subsequent R peaks are referred to as RR intervals. The electrochemical process that gives rise to each R peak in the heart has been shown to be modeled by a Gaussian random walk with linear drift \cite{barbieri2005point}, with the RR intervals therefore following an inverse Gaussian distribution \cite{barbieri2005point, chhikara1988inverse}. By rigorously modeling the generative process of a heartbeat, the resulting model allows for a degree of accuracy and precision in tracking subtle but physiologically important beat-to-beat variations not achieved by other standard smoothing or averaging techniques. The lognormal distribution has also been employed to model physiological time series (including heartbeat dynamics in children \cite{gee2016predicting}) because of its similarity to the inverse Gaussian as a heavier-tailed distribution \cite{subramanian2020point}. Therefore, the lognormal neural TPP is an apt initial choice of neural model for heartbeat dynamics applications.
\vspace*{-.0cm} \section{Methods and Results} 
\begin{figure}
    \centering \vspace*{-.1cm}
    \includegraphics[width=0.8\textwidth]{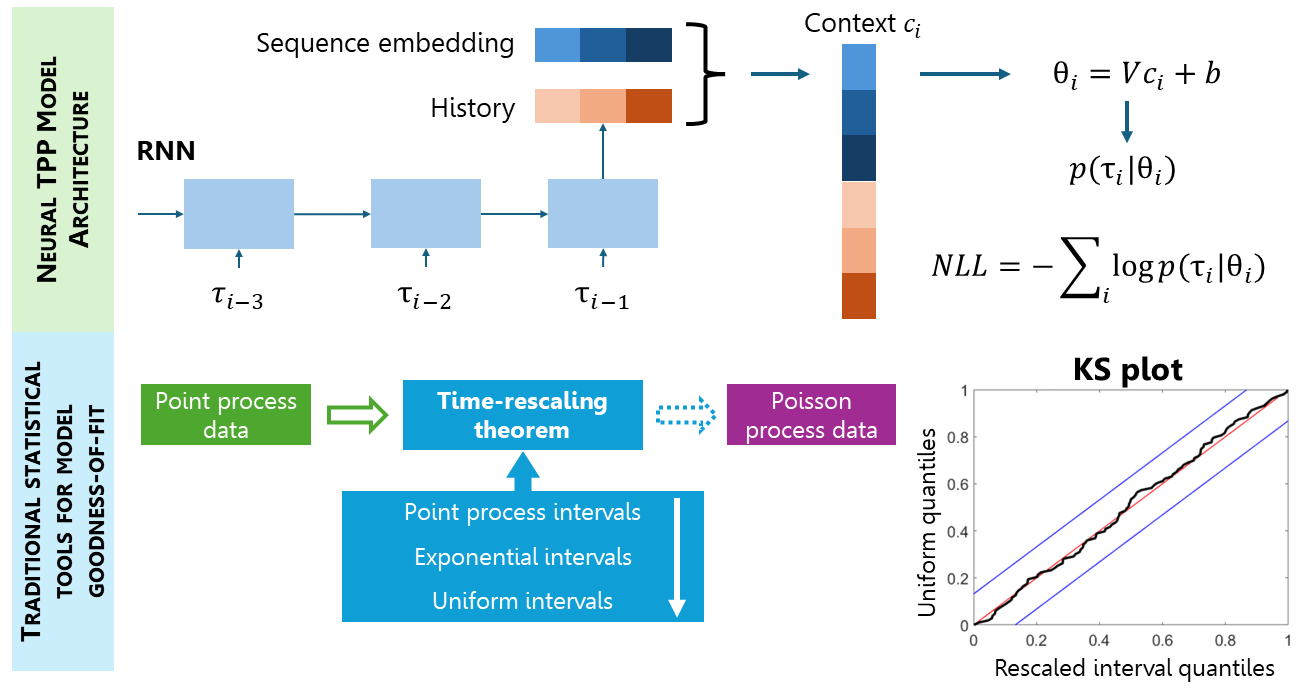}\vspace*{-.1cm}
    \caption{(Top) An architecture diagram for the density-based neural TPP model. (Bottom) The goodness-of-fit (GOF) framework. The time-rescaling theorem is used to test the GOF of the density neural TPP model with standard statistical methodology like the KS-plot.} \vspace*{-.2cm}
    \label{fig:hero}
\end{figure}

\vspace*{-.1cm}
\label{headings}
Fig~\ref{fig:hero} illustrates the Neural TPP architecture and the GOF framework that we leverage in this work to evaluate Neural TPP models. Using this GOF framework, we compute the the KS-distance as the primary measure of model performance. We consider the publicly available MIT-BIH normal sinus rhythm dataset\footnote{\label{note:url}\url{https://www.physionet.org/content/nsrdb/1.0.0/
}}. This dataset consists of 130 minutes of ECG recording of normal sinus rhythm from each of 18 subjects (ages 20-50, 13 women). Each subject's ECG data was preprocessed using the Pan-Tompkins algorithm \cite{pan1985real} to extract the times of the R peaks. 
\vspace*{-.0cm} \subsection{Hyperparameter Exploration} \vspace*{-.1cm}\label{sec:hyper}
We split each subject's data into sequences of R peak times in non-overlapping 10-minute segments. The first 11 of those sequences per subject were for training (191 sequences total), the next one was for validation (18 sequences total), and the last for the test set (18 sequences total). We explored modifying context length $C$ (16, 32, 64, 128), the number of components $K$ (1, 2, 4, 8, 16), and type of RNN (GRU, RNN, LSTM). We used the validation set for early stopping. We computed the 95\% confidence interval (CI) for the mean KS-distance across the test set along with the mean 95\% significance cutoff for the KS-distance for each run. Full details are available in Appendix~\ref{app:hyperopt}, Table~\ref{tab:hyperparameter-exp-results}.

We found that the model is most sensitive to RNN type, with GRU consistently outperforming the other types. Increasing model size ($C$, $K$) improves performance up to a certain point ($C = 64$ and $K = 8$) after which there are diminishing returns. 
\vspace*{-.1cm}\subsection{Varying Training Sequence Length}\vspace*{-.1cm}
We sought to evaluate the sensitivity of the model to shorter training sequences while keeping the test sequence length constant. We divided the 110 minutes of training data for each subject into different length sequences of 1, 2, or 5 minutes, while keeping the same 10-minute validation and test set sequences. For each training sequence length, we varied hyperparameters and measured model performance as in Section~\ref{sec:hyper}. Full details are available in Appendix~\ref{app:trainlen}, Table~\ref{tab:train-seq-length-results}. We find the best model performance for training sequences around 2-5 minutes, with decreased performance for training sequences both shorter and longer than that (Fig~\ref{fig:train-seq}). 
\vspace{-.1cm} \subsection{Zero-shot Experiments} \vspace{-.1cm}
We sought to evaluate the zero-shot predictive power of the density-based neural TPP framework on longer time-frame data from new subjects (as would be admitted to a hospital). Therefore, we trained the model on data from 17 subjects for $1000$ epochs using $5$-minute training sequences and evaluated the trained model on both 10-minute and 30-minute sequences from a held-out test subject. We rotated each subject as the test subject. We computed the 95\% CI for the mean KS-distance across test subject sequences along with the mean 95\% significance cutoff for the KS-distance.

\begin{figure}
    \centering
    \hspace*{-1.5cm} \vspace*{-.2cm}\includegraphics[width=1.2\textwidth]{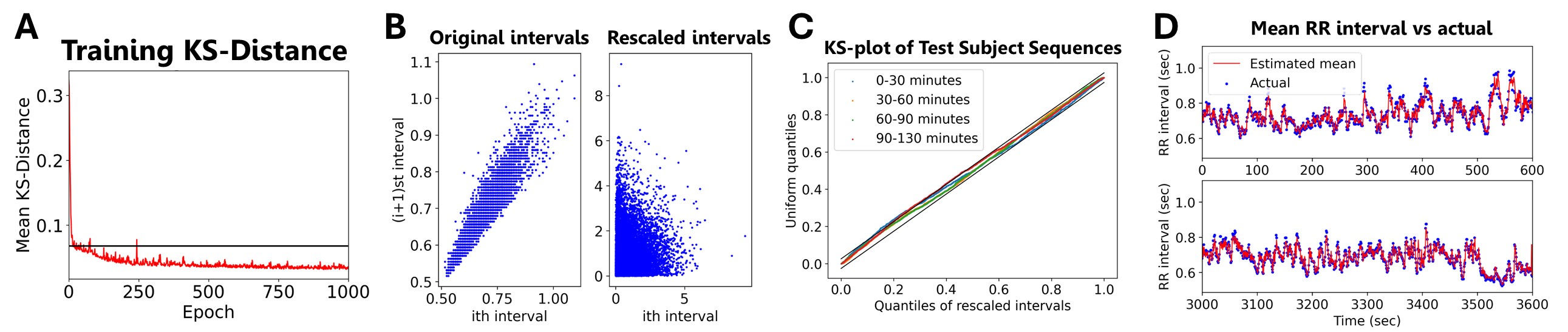}
    \caption{Zero-shot prediction results for one test subject with 30-minute test sequences. (a) Training fit improves dramatically with more training, with mean KS-distance (red) well below cutoff (black). (b) Rescaling original RR intervals from the test subject using the time-rescaling theorem yields uncorrelated rescaled intervals, as expected for a well-fitting model. (c) The KS-plot shows remarkable fit across all 30-minute test subject sequences, with rescaled interval quantiles aligned closely with the $y=x$ line and falling almost fully within the significance cutoffs (black). (d) The trained model tightly predicts the mean RR interval using a mixture of lognormal densities.} \vspace*{-.3cm}
    \label{fig:zero-shot-1}
\end{figure}

Fig~\ref{fig:zero-shot-1} and Fig~\ref{fig:zero-shot-2} (Appendix~\ref{app:zero-shot}) show the zero-shot prediction results for 2 example subjects. The 95\% CI for the test subject mean KS-distance was fully below the mean significance cutoff for 13 subjects with 10-minute test sequences and 5 subjects with 30-minute test sequences. The 95\% CI contained the mean significance cutoff for the remaining 5 subjects with 10-minute test sequences and 9 of the remaining subjects with 30-minute test sequences. Results are summarized in Appendix~\ref{app:zero-shot}, Table~\ref{tab:zero-summary}. Unsurprisingly, the zero-shot prediction is better for shorter length sequences. However, as Figs~\ref{fig:zero-shot-1} and \ref{fig:zero-shot-2} demonstrate, the zero-shot prediction performance is still remarkable on longer test sequences. 

\vspace{-.1cm} \section{Discussion} \vspace{-.1cm}
In this work, we present three principal findings. First, we successfully adapted classical GOF frameworks to evaluate neural TPP model fit on per dataset basis. This opens the door for statistically rigorous evaluation of neural TPPs and for systematic comparisons with existing point process methods. We will do this comparison in future work. We will also test this framework on more and larger ECG datasets, extend to other physiological time series, and explore modeling multi-sensor physiological time series.

Second, with respect to heartbeat dynamics in normal sinus rhythm, the ability of the neural TPP model to capture nonlinearity and long-range temporal dependencies through model complexity allows us to measure how much `memory' the physiological phenomena requires. In this case, the system has memory, evidenced by the sensitivity of the model to the specific RNN architecture (GRU vs others) and the effectiveness of 2-5 minute vs 1-minute training sequences. However, this memory is not infinite, as evidenced by the diminishing returns on performance with longer training sequences or continually increasing model size. This result supports previous work in the field, where limited range history-dependent models have been successfully employed to represent the known influence of respiration and basic hemodynamic reflexes \cite{barbieri2005point, gee2016predicting}. The required 'memory' of the system may change with specific pathology, however.  

Third, the framework of density-based neural TPPs retains the rich physiologically relevant point process prior, but also allows for shared learnable parameters. Our experiments with zero-shot prediction revealed that neural TPPs can learn generalizable models of human heartbeat dynamics during normal sinus rhythm for new subjects. In future work, we will probe this generalizability in the setting of different pathologies.

This work represents an important first step towards bridging the fields of physiology, statistics, and AI to yield generalizable models of physiological processes.


\subsubsection*{Acknowledgments}
S.S. is funded by Schmidt Science Fellows and the L'Oreal USA FWIS fellowship.

\bibliography{iclr2024_conference}
\bibliographystyle{iclr2024_conference}

\appendix
\section{Appendix}

\subsection{Point Processes}
\label{app:point}

We define a temporal point process as a random process which generates a series of arrival times $0 < t_1 < \dotsc < t_n$. We define $\tau_i = t_i - t_{i-1}$ to be the interevent time. We define the set of events $\mathcal{T} = \{t_1,\dotsc,t_n\}$. The next arrival time may depend on the history of the process thus far defined as $\mathcal{H}_t = \{t_j \in \mathcal{T} | t_j < t\}$.

A point process is uniquely defined by its conditional intensity function $\lambda^*_\theta(t) = \lambda(t|\mathcal{H}_t)$. The conditional intensity function is defined as the probability of an event occurring in an infinitesimal time interval
\begin{align}
    \lambda^*_\theta(t) &= \lim_{\Delta \to 0} \frac{P(\mathrm{event\ in\ }[t, t+\Delta)| \mathcal{H}_t)}{\Delta}
\end{align}
We will also find it useful to alternatively consider the conditional density function $p^*_\theta(\tau_i) = p_\theta(\tau_i | \mathcal{H}_t)$ and conditional cumulative density function $F^*_\theta$. The density and intensity functions are related by the following equations
\begin{align}
    p^*_\theta(\tau_i)  
    = \lambda^*_\theta(t_{i-1}+\tau_i)\exp\left(-\int_{0}^{\tau_i} \lambda^*_\theta(t_{i-1}+s) ds \right ), \quad
    \lambda^*_\theta(\tau_i) = \frac{p^*_\theta(\tau_i)}{1-F^*_\theta(\tau_i)}
\end{align}
The log-likelihood $\mathcal{L}$ of a series of events is given by
\begin{align}
    \mathcal{L} &= \sum_i \log p^*_\theta(\tau_i) 
    = \sum_i \log \lambda^*_\theta(t_i) - \int_0^{t_N} \lambda^*_\theta(s) ds
\end{align}

\subsection{Optimization and Hyperparameters}
\label{app:hyperopt}
We performed systematic hyperparameter optimization of neural TPP parameters including the context size $C$, the number of mixture components $K$, the choice of RNN cell, and the number of training epochs. All sequences are of length 10 minutes. The 130 minutes of ECG data per each subject are chopped into 13 10-minute sequences of R peak times. These sequences are split across training, validation and test. 110 minutes of data are used for training, 10 for validation and 10 for test. Models were optimized using Adam \cite{kingma2014adam} on a T4 GPU. Training runs typically took under 10 minutes. Full results are provided in Table~\ref{tab:hyperparameter-exp-results}.

\begin{table}[ht]
\caption{Hyperparameter Exploration. $C$ is the context length. $K$ is the number of components. Epochs is the maximum number of epochs allowed (but models may stop earlier due to early stopping.) RNN indicates type of RNN cell. All training sequences are of length 10 minutes. 95\% CI Test KS indicates the 95\% confidence interval for the mean KS-distance on the test set. 95\% CI Val KS and 95\% CI Train KS have similar meanings. The mean 95\% test cutoff KS-distance is $0.0471$. The mean 95\% validation cutoff KS-distance is $0.0477$. The mean 95\% train cutoff KS-distance is $0.0481$.}
\label{tab:hyperparameter-exp-results}
\centering
\resizebox{\columnwidth}{!}{
\begin{tabular}{rrlrlll}
\cline{1-7}
\multicolumn{1}{l}{$C$} & \multicolumn{1}{l}{$K$} & \textbf{RNN} & \multicolumn{1}{l}{\textbf{Epochs}} & \textbf{95\% CI Test KS} & \textbf{95\% CI Val KS} & \textbf{95\% CI Train KS} \\
\cline{1-7}
64 & 8 & GRU & 2000 & 0.0225, 0.0314 & 0.0346, 0.0829 & 0.0501, 0.0636\\
128 & 8 & GRU & 1500 & 0.0229, 0.0307 & 0.0228, 0.0323 & 0.0277, 0.0309 \\
32 & 8 & GRU & 500 & 0.0332, 0.0434 & 0.0353, 0.0474 & 0.0401, 0.0453 \\
64 & 8 & RNN & 500 & 0.0396, 0.0514 & 0.0421, 0.0626 & 0.0493, 0.0562 \\
64 & 8 & LSTM & 500 & 0.0377, 0.0538 & 0.0394, 0.0575 & 0.0436, 0.0499 \\
64 & 16 & GRU & 500 & 0.0322, 0.0417 & 0.0302, 0.0411 & 0.0344, 0.0392 \\
64 & 32 & GRU & 1000 & 0.0258, 0.0331 & 0.0236, 0.0331 & 0.0265, 0.0293 \\
64 & 4 & GRU & 500 & 0.028, 0.0389 & 0.0361, 0.0466 & 0.0402, 0.045 \\
64 & 2 & GRU & 1000 & 0.0282, 0.0383 & 0.0273, 0.0415 & 0.0328, 0.0368 \\
64 & 1 & GRU & 500 & 0.0405, 0.0577 & 0.0363, 0.0587 & 0.0491, 0.0563 \\
128 & 32 & GRU & 500 & 0.0285, 0.0402 & 0.0301, 0.0446 & 0.0328, 0.0373 \\
128 & 2 & GRU & 500 & 0.0314,  0.0399 & 0.0304, 0.0428 & 0.0341, 0.0382 \\
\end{tabular}
}
\end{table}

\subsection{Training Sequence Lengths}
\label{app:trainlen}
\begin{figure}
    \centering
    \includegraphics[width=0.6\textwidth]{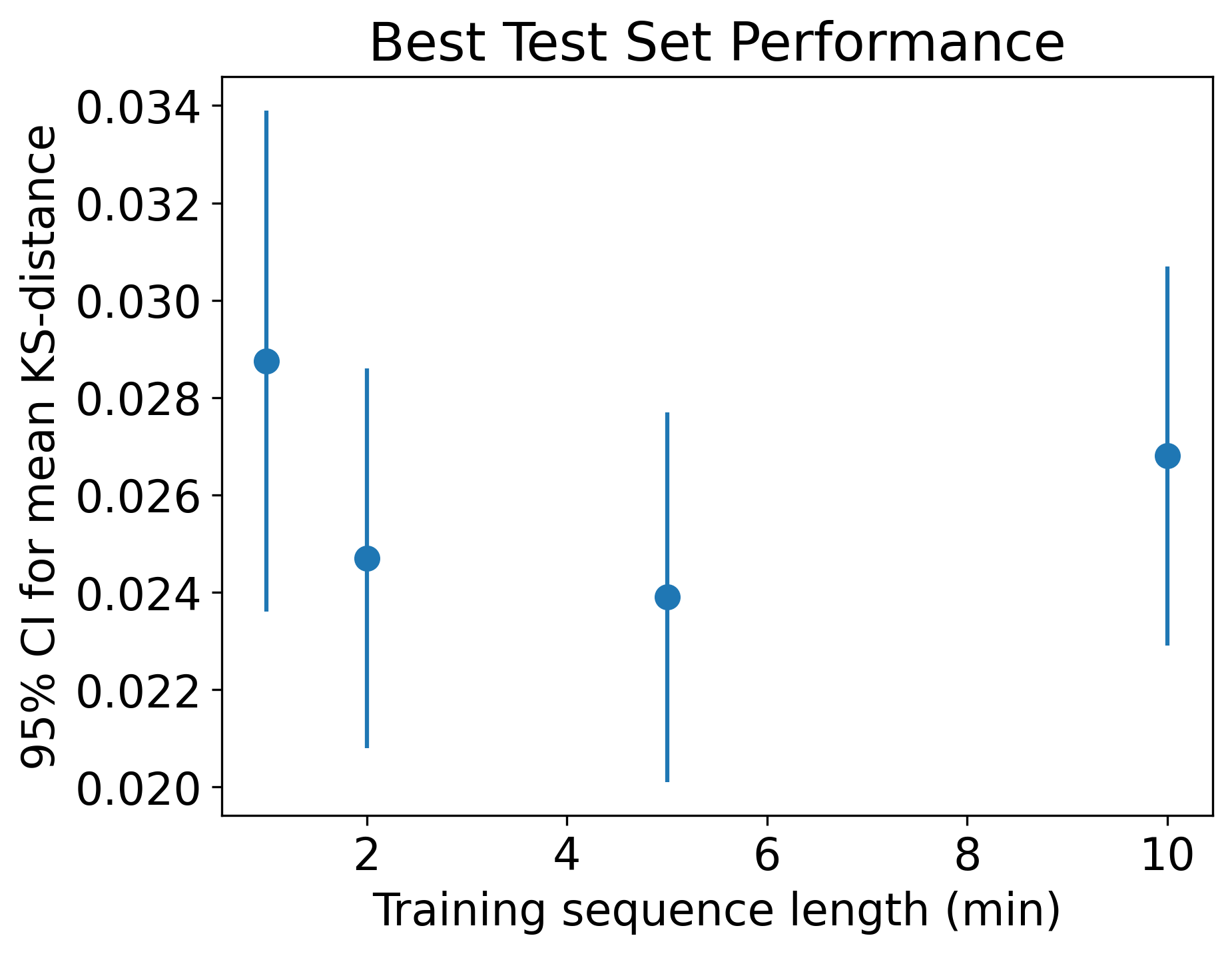}
    \caption{The 95\% confidence intervals for the mean KS-distances of the best models on the test set with varying training sequence lengths. Between 2-5 minutes of training seems to generalize best.}
    \label{fig:train-seq}
\end{figure}

We followed the same general procedure as in Section~\ref{app:hyperopt}, but divided the 110 minutes of training data for each subject divided into different length sequences of 1,2,5 or 10 minutes. The validation set still contained a single 10-minute sequence for each subject and the test set was still a single 10-minute sequence for each subject. See full results in Table~\ref{tab:train-seq-length-results}. Fig~\ref{fig:train-seq} summarizes results and shows that between 2-5 minutes of training sequence length is sufficient for best results.
\begin{table}[]
\caption{Training Sequence Length Variation.  $C$ is the context length. $K$ is the number of components. $L$ is the length of the training sequences in minutes. Epochs is the maximum number of epochs allowed (but models may stop earlier due to early stopping.) RNN indicates type of RNN cell. 95\% CI Test KS indicates the 95\% confidence interval for the mean KS-distance on the test set. 95\% CI Val KS and 95\% CI Train KS have similar meanings. The mean 95\% test cutoff KS-distance is $0.0471$. The mean validation cutoff KS-distance is $0.0477$. The mean 95\% train cutoff KS-distance is $0.0681$ for 5-minute sequences,  $0.1082$ for 2-minute sequences, $0.1541$ for 1-minute sequences.}
\label{tab:train-seq-length-results}
\centering
\resizebox{\columnwidth}{!}{
\begin{tabular}{rrllrlll}
\cline{1-8}
\multicolumn{1}{l}{$C$} & \multicolumn{1}{l}{$K$} & \textbf{RNN} & $L$ & \multicolumn{1}{l}{\textbf{Epochs}} & \textbf{95\% CI Test KS} & \textbf{95\% CI Val KS} & \textbf{95\% CI Train KS} \\
\cline{1-8}
64 & 8 & GRU & 5 min & 2000 & 0.0241, 0.0381 & 0.0186, 0.0282 & 0.0315, 0.0336 \\
128 & 8 & GRU & 5 min & 1000 & 0.0225, 0.0322 & 0.0242, 0.0393 & 0.0355, 0.0384 \\
32 & 8 & GRU & 5 min & 3000 & 0.0227, 0.0306 & 0.022, 0.0294 & 0.0321, 0.0344 \\
16 & 8 & GRU & 5 min & 1000 & 0.0272, 0.0352 & 0.029, 0.0369 & 0.0368, 0.0393 \\
32 & 8 & RNN & 5 min & 1000 & 0.0338, 0.0432 & 0.0306, 0.0423 & 0.0417, 0.0455 \\
32 & 8 & LSTM & 5 min & 1000 & 0.0237, 0.0313 & 0.0227, 0.035 & 0.0357, 0.0384 \\
32 & 16 & GRU & 5 min & 2000 & 0.023, 0.0307 & 0.0225, 0.0335 & 0.0326, 0.0348 \\
32 & 4 & GRU & 5 min & 1000 & 0.0247, 0.0346 & 0.0279, 0.0447 & 0.0404, 0.0439 \\
32 & 2 & GRU & 5 min & 1000 & 0.0293, 0.0386 & 0.0273, 0.041 & 0.04, 0.0433 \\
32 & 1 & GRU & 5 min & 1000 & 0.0349, 0.0514 & 0.0323, 0.0515 & 0.0503, 0.0545 \\
128 & 16 & GRU & 5 min & 2000 & 0.0201, 0.0288 & 0.0211, 0.0354 & 0.0319, 0.0342 \\
64 & 16 & GRU & 5 min & 2000 & 0.0201, 0.0277 & 0.0218, 0.0296 & 0.0316, 0.0337 \\
64 & 8 & GRU & 2 min & 1000 & 0.0208, 0.0286 & 0.023, 0.0319 & 0.0471, 0.0491 \\
64 & 8 & RNN & 2 min & 1000 & 0.0256, 0.0348 & 0.0277, 0.0398 & 0.0537, 0.0561 \\
64 & 8 & LSTM & 2 min & 1000 & 0.0233, 0.0313 & 0.0242, 0.0321 & 0.0531, 0.0555 \\
128 & 8 & GRU & 2 min & 1000 & 0.0267, 0.0348 & 0.0199, 0.0295 & 0.0466, 0.0486 \\
32 & 8 & GRU & 2 min & 2000 & 0.0219, 0.031 & 0.0224, 0.034 & 0.0481, 0.0502 \\
16 & 8 & GRU & 2 min & 3000 & 0.0238, 0.0333 & 0.0254, 0.0347 & 0.0501, 0.0523 \\
64 & 16 & GRU & 2 min & 1000 & 0.0262, 0.035 & 0.0223, 0.0322 & 0.0471, 0.049 \\
64 & 4 & GRU & 2 min & 1000 & 0.0255, 0.0373 & 0.0245, 0.0359 & 0.0486, 0.0507 \\
64 & 2 & GRU & 2 min & 1000 & 0.0286, 0.0417 & 0.0281, 0.043 & 0.053, 0.0553 \\
64 & 1 & GRU & 2 min & 1000 & 0.0328, 0.052 & 0.0314, 0.0472 & 0.0578, 0.0606 \\
64 & 8 & GRU & 1 min & 1000 & 0.0256, 0.0369 & 0.0281, 0.0411 & 0.0679, 0.0701 \\
128 & 8 & GRU & 1 min & 1000 & 0.0283, 0.039 & 0.0327, 0.0439 & 0.0685, 0.0709 \\
32 & 8 & GRU & 1 min & 1000 & 0.0258, 0.0381 & 0.0358, 0.0533 & 0.0734, 0.0758 \\
16 & 8 & GRU & 1 min & 2000 & 0.0236, 0.0339 & 0.0242, 0.0343 & 0.0647, 0.0667 \\
16 & 8 & RNN & 1 min & 1500 & 0.0291, 0.0406 & 0.0359, 0.0488 & 0.0691, 0.0713 \\
16 & 8 & LSTM & 1 min & 1500 & 0.0251, 0.0357 & 0.0261, 0.0349 & 0.0647, 0.0667 \\
16 & 16 & GRU & 1 min & 1500 & 0.0257, 0.0342 & 0.0236, 0.0332 & 0.0666, 0.0687 \\
16 & 4 & GRU & 1 min & 1500 & 0.0259, 0.0348 & 0.0244, 0.0368 & 0.0681, 0.0704 \\
16 & 2 & GRU & 1 min & 1500 & 0.0284, 0.0397 & 0.0258, 0.0378 & 0.0692, 0.0715 \\
16 & 1 & GRU & 1 min & 1500 & 0.033, 0.046 & 0.0319, 0.0481 & 0.0768, 0.0793
\end{tabular}
}
\end{table}

\subsection{Zero-shot Experiments}
\label{app:zero-shot}
\begin{figure} 
    \centering
    \hspace*{-1.5cm}
    \includegraphics[width=1.2\textwidth]{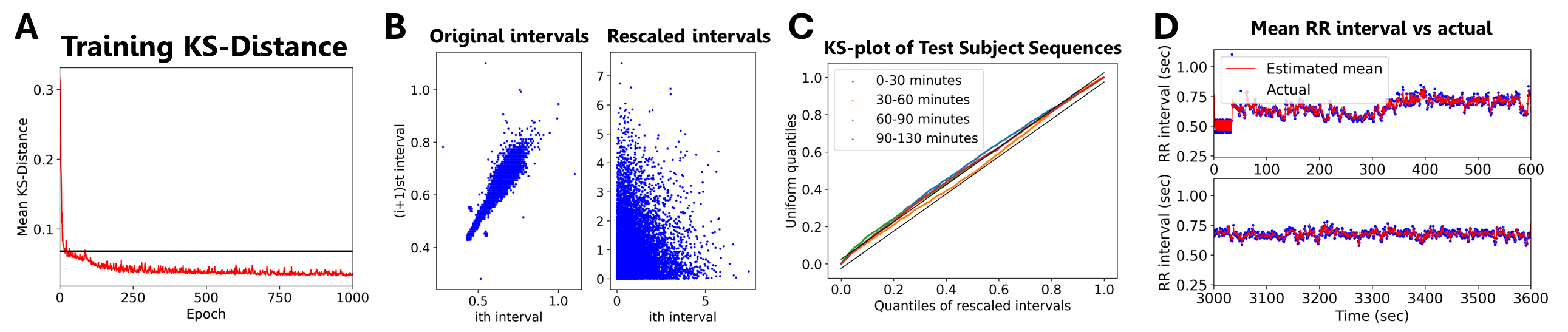}
    \caption{Zero-shot prediction results for a different subject than Fig~\ref{fig:zero-shot-1} with 30-minute test sequences. (a) Training data fit improves dramatically with more epochs of training, with the mean KS-distance (red) well below cutoff (black). (b) Rescaling the original RR intervals from the test subject using the time-rescaling theorem yields uncorrelated rescaled intervals, as would be expected for a well-fitting model. (c)  The KS-plot shows reasonable fit across all 30-minute test subject sequences, but with more variation than Fig~\ref{fig:zero-shot-1}. (d) The trained model tightly predicts the mean RR interval using a mixture of lognormal densities.} 
    \label{fig:zero-shot-2}
\end{figure}

Fig~\ref{fig:zero-shot-2} provides full zero-shot visualizations for an additional subject. Full details of zero-shot experiments are provided in Table~\ref{zero-shot-results}. Table~\ref{tab:zero-summary} summarizes results across all subjects.  
\begin{table}[]
\caption{Zero-Shot Testing. Subject indicates the subject used as the test subject. $L_{train}$ is the length of the training sequences in minutes. $L_{test}$ is the length of the test sequences in minutes. All models were trained for 1000 epochs. RNN indicates type of RNN cell. 95\% Test KS indicates the 95\% confidence interval for the mean KS-distance on the test set. 95\% Val KS and 95\% CI Train KS have similar meanings. Test cut indicates the mean 95\% cutoff for the test set KS-distance. Train cut indicates the mean 95\% cutoff for the training set KS-distance.}
\label{zero-shot-results}
\centering
\resizebox{\columnwidth}{!}{
\begin{tabular}{rlllrlr}
\cline{1-7}
\multicolumn{1}{l}{\textbf{Subject}} & $L_{train}$ & $L_{test}$ & \textbf{95\% Test KS} & \multicolumn{1}{l}{\textbf{Test Cut.}} & \textbf{95\% Train KS} & \multicolumn{1}{l}{\textbf{Train Cut.}} \\
\cline{1-7}
1 & 5 min & 10 min & 0.0224, 0.0329 & 0.0459 & 0.0375, 0.0405 & 0.0681 \\
2 & 5 min & 10 min & 0.039, 0.0537 & 0.055 & 0.0337, 0.0359 & 0.0674 \\
3 & 5 min & 10 min & 0.0271, 0.0394 & 0.0482 & 0.0336, 0.0357 & 0.0679 \\
4 & 5 min & 10 min & 0.0332,  0.0501 & 0.0476 & 0.0326, 0.0346 & 0.068 \\
5 & 5 min & 10 min & 0.042, 0.0509 & 0.0445 & 0.0336, 0.0356 & 0.0682 \\
6 & 5 min & 10 min & 0.0445, 0.0755 & 0.0519 & 0.0329, 0.0349 & 0.0676 \\
7 & 5 min & 10 min & 0.0184, 0.0302 & 0.05 & 0.0366, 0.0391 & 0.0678 \\
8 & 5 min & 10 min & 0.026, 0.0347 & 0.0504 & 0.0331, 0.0353 & 0.0678 \\
9 & 5 min & 10 min & 0.028, 0.0462 & 0.0487 & 0.0334, 0.0357 & 0.0679 \\
10 & 5 min & 10 min & 0.0304, 0.0584 & 0.0525 & 0.0343, 0.0368 & 0.0676 \\
11 & 5 min & 10 min & 0.0206, 0.0307 & 0.0463 & 0.0325, 0.0348 & 0.0681 \\
12 & 5 min & 10 min & 0.0235, 0.0315 & 0.0451 & 0.0349, 0.0373 & 0.0682 \\
13 & 5 min & 10 min & 0.019, 0.0238 & 0.0471 & 0.0324, 0.0344 & 0.068 \\
14 & 5 min & 10 min & 0.0272, 0.0507 & 0.0442 & 0.0358, 0.0389 & 0.0683 \\
15 & 5 min & 10 min & 0.0331, 0.046 & 0.0479 & 0.0351, 0.0374 & 0.068 \\
16 & 5 min & 10 min & 0.0204, 0.0309 & 0.0514 & 0.0321, 0.0342 & 0.0677 \\
17 & 5 min & 10 min & 0.0189, 0.0262 & 0.0462 & 0.0326, 0.0346 & 0.0681 \\
18 & 5 min & 10 min & 0.0231, 0.0388 & 0.0403 & 0.0339, 0.0363 & 0.0686 \\
1 & 5 min & 30 min & 0.0399, 0.0444 & 0.0256 & 0.0403, 0.0433 & 0.0681 \\
2 & 5 min & 30 min & 0.0127, 0.0624 & 0.0307 & 0.0337, 0.0363 & 0.0674 \\
3 & 5 min & 30 min & 0.0203, 0.0283 & 0.0269 & 0.0326, 0.0348 & 0.0679 \\
4 & 5 min & 30 min & 0.0194, 0.0319 & 0.0265 & 0.0323, 0.0344 & 0.068 \\
5 & 5 min & 30 min & 0.0377, 0.0544 & 0.0248 & 0.0342, 0.0365 & 0.0682 \\
6 & 5 min & 30 min & 0.0272, 0.0774 & 0.0287 & 0.0322, 0.0344 & 0.0676 \\
7 & 5 min & 30 min & 0.0153, 0.0182 & 0.0278 & 0.0371, 0.0394 & 0.0678 \\
8 & 5 min & 30 min & 0.0219, 0.0333 & 0.028 & 0.0338, 0.0363 & 0.0678 \\
9 & 5 min & 30 min & 0.0075, 0.0584 & 0.0274 & 0.0394, 0.0423 & 0.0679 \\
10 & 5 min & 30 min & 0.0231, 0.0519 & 0.0292 & 0.0332, 0.0354 & 0.0676 \\
11 & 5 min & 30 min & 0.0134, 0.0265 & 0.0259 & 0.0347, 0.0372 & 0.0681 \\
12 & 5 min & 30 min & 0.0424, 0.052 & 0.025 & 0.0363, 0.0388 & 0.0682 \\
13 & 5 min & 30 min & 0.0141, 0.0203 & 0.0262 & 0.0346, 0.037 & 0.068 \\
14 & 5 min & 30 min & 0.02, 0.0407 & 0.0246 & 0.0349, 0.0374 & 0.0683 \\
15 & 5 min & 30 min & 0.0129, 0.0208 & 0.0266 & 0.0412, 0.0444 & 0.068 \\
16 & 5 min & 30 min & 0.01, 0.0234 & 0.0287 & 0.0356, 0.0381 & 0.0677 \\
17 & 5 min & 30 min & 0.0129, 0.0226 & 0.0256 & 0.0345, 0.0369 & 0.0681 \\
18 & 5 min & 30 min & 0.0542, 0.1051 & 0.0225 & 0.038, 0.0412 & 0.0686
\end{tabular}
}
\end{table}

\begin{table}[] 
\caption{Zero-shot results summary. $L_{test}$ is the length of the test sequence. $N_{under}$ is the number of test subjects fully under KS-cutoff, $N_{overlap}$ is the number where the KS-cutoff is within the 95\%-CI for the subject, $N_{over}$ is the number where the 95\%-CI is over the KS-cutoff,  and $N_{total}$ is the total number of subjects\\}
\label{tab:zero-summary}
\centering
\begin{tabular}{rrrrl}
\cline{1-5}
$L_{test}$ & $N_{under}$ & $N_{overlap}$ & $N_{over}$ &  $N_{total}$  \\
\cline{1-5}
10 & 13 & 5 & 0 & \textbf{18}\\
30 & 5 & 9 & 4 & \textbf{18}
\end{tabular}
\end{table}

\end{document}